\documentclass[12pt]{article}
\usepackage{epsf}

\begin{document}

\title{Instantons in the Langevin dynamics: an application to spin glasses.}
\author{A. V. Lopatin, L. B. Ioffe}
\date{\today}
\maketitle

\begin{abstract}
We develop a general technique to calculate the probability of transitions
over the barriers in spin-glasses in the framework of the dynamical theory.
We use Lagrangian formulation of the instanton dynamics in which the
transitions are represented by instantons. We derive the full set of the
equations that determine the instantons but instead of solving them directly
we prove that an instanton process can be mapped into a usual process going
back in time which simplifies the problem significantly. We apply this
general considerations to a simple example of the spherical
Sherrington-Kirkpatrick model and we find the probability of the transition
between the metastable states which is in agreement with physical
expectations.
\end{abstract}

\section{Introduction}

It is well known that all spin-glass systems have complicated free energy
structures with many metastable states separated by large barriers \cite
{Fischer93}. The calculation of heights of these barriers is a difficult
task even for the long-range spin-glass models because most of the methods
developed in the spin glass theory do not give such information. For
example, the application of the standard replica method \cite{Fischer93}
allows to find the equilibrium free energy which does not contain any
information about the heights of the barriers. The standard dynamical
approach \cite{Dotsenko90,kucu} gives the properties of different
metastable states and their history dependence and might contain, in
principle, the information about the transition rates between them but in
the limit of long-range interaction the probability of transitions becomes
exponentially low in the number of spins, $N$, and, thus, such processes are
neglected by the mean field derivation of the equations for the correlation
and response functions on which this aproach is usually based. The
modification of the replica method that allows to estimate the energy
barriers between metastable states was suggested in Ref.\cite{cgp}. The main
idea of this modification is to study the free energy of the state
constrained to have a certain overlap with the given state. The main
drawback of it is that it is not clear whether the state corresponding to
the energy maximum found by this method is dynamically accessible and that
it is indeed a bottleneck of a transition process. The problem is
exacerbated by the fact that the replica method weights all states with the
Boltzmann weight so it might miss the rare saddle points of low energy in
favour of more abundant high energy ones. In this paper we develop an
alternative technique which is based on the modified dynamical approach \cite
{Ioffe98} for the calculation of the barriers between the metastable states. 

Before we discuss how to modify the dynamical theory so it does not neglect
rare processes such as transitions over the barriers we briefly review the
standard dynamical approach to the spin-glasses. In this approach one starts
from the Lagrangian formulation of the Langevin dynamics, averages over the
disorder and, making the saddle point approximation, arrives at a closed
system of equations for the sigle site spin-spin correlation function $%
D(t_{1},t_{2})$ and response function $G(t_{1},t_{2})$. The corrections to
the saddle point approximation are small in $1/N$ because interaction has a
range $N$. For example, in the case of the spherical p-spin interacting
model one gets a set of integro-differential equations which were solved
numerically (and partially analytically); the solution is made possible by
the fact that these equations are \emph{forward propagating} in time which
is in turn due to the causality of the Langevin equations and initial
conditions imposed in the past \cite{kucu,pspin}. By construction these mean
field equations describe the most probable evolution of the system in time
and ignore all rare processes such as transitions over the barriers.
Empirically we visualize the processes that give the main contribution to
the conventional dynamical theory as motion down the energy landscape or
small fluctuations near the bottom of the valley.

Consider now a typical dynamical process that corresponds to the transition
between two close metastable states in a free energy landscape shown in
Fig.~1. The system initially is in the state 1 and we want to find the
probability of the transition to the state 2. The path from point 1 to point
2 consists of the uphill motion from state 1 to the unstable stationary
point 3 and the motion from unstable point 3 to stable state 2. Only the
first part of the motion corresponds to the rare process, and, therefore,
the probability of the transition between the states 1 and 2 is determined
by the uphill motion from 1 to 3. In the Lagrangian formulation of the
Langevin dynamics the uphill motion can be described as an instanton which
action gives the probability of the process. Note that to find this solution
one needs to ``force'' the system to go upward, i.e. one needs to apply a
boundary condition in future (at point 3 in our example) that destroys
causality of the theory. This complicates enormously the dynamical equations
describing the instanton motion compared to the dynamical equations
describing typical processes (such as motion from 3 to 2).

The main result of this paper is that an instanton motion can be
mapped into a usual motion going back in time. In  our example it means that
the uphill motion from point 1 to point 3 can be mapped into the usual
downhill motion from point 3 to point 1. This allows one to solve the usual 
\emph{forward propagating} equations instead of solving the complicated
equations describing the instanton motion. In general, this method does not
allow one to find the barrier beween one given state and another but it
allows one to find some instanton processes which, hopefully, correspond to
typical barriers in a system. In the simple example of a spherical
Sherrington-Kirkpatrick model which we consider here it gives the barrier
that separates doubly degenerate ground states that differ by the sign of
the magnetization; this barrier is physically important because it controls
the decay of the ground state magnetization in this system. In this model
there are only two locally stable states and $N$ locally unstable ones,
naturally one expects that these unstable states are saddle points and the
trajectory which connects two ground states must go through one of them but
these general qualitative arguments do not indicate which of these saddle
points should be used. Qualitatively the problem is that even in this simple
model the lowest saddle point might not connect two different minima but
connect one minima to itself (false pass). Our approach proves that it is
sufficient to climb up to the lowest of them in order to get from one ground
state to another.

\begin{figure}[h]
\unitlength1.0cm
\par
\begin{center}
\begin{picture}(6,2.4)
\epsfxsize=6.0 cm
\epsfysize=2.4 cm
\epsfbox{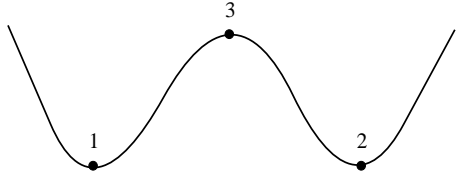}
\end{picture}
\end{center}
\caption{Energy landscape illustrating a transition between states 1 and 2
via saddle point 3. Uphill motion from 1 to 3 is a rare process which is
missed by conventional mean field dynamical equations for the spin glasses,
the action corresponding to this process is large in $N$. Downhill motion
from 3 to 2 does not cost any action and is described by the conventional
dynamic equations, its action is zero.}
\end{figure}

Further, we confirm that in case when the free energy landscape of the
problem is known explicitly, the probability of the transition obtained as
an action of the instanton solution is $e^{-\Delta F/T}$ where $\Delta F$ is
the difference between the free energy at the end of the instanton
trajectory (point 3) and the free energy of the stationary state (point 1).
The free energy at the unstable fixed point 3 should be understood as $%
F=E-TS,$ where $E$ is the energy and $S$ is the entropy defined as a
logarithm of the configuration space restricted to the direction
perpendicular to the trajectory. Of course, there is no much need to calculate 
the transition probability if the energy of the saddle point in known exactly and
it is established that the saddle point is dynamically accessible; the
advantage of the method is that it can be also used in the cases where energy
lanscape can not be found explicitly, e.g. in all problems in which the
disorder average was performed first.

The paper is organized as follows: In Section \ref{2} we prove that an
instanton equation of motion can be mapped into a usual equation of motion
reversed in time. In Section \ref{3} we consider an instanton transition in
the spherical SK model. Section \ref{4} summarizes our results.

\section{Instanton equations}

\label{2}

We start from the Langevin equation describing the overdamped relaxation of
the system with energy $\mathcal{H}$

\[
\Gamma _{0}^{-1}\partial _{t}\sigma _{i}=-{\frac{{\delta \beta \mathcal{H}}}{%
{\delta \sigma _{i}}}}+\xi _{i}. 
\]
Here $\xi _{i}$ is the Langevin noise with the correlator 
\begin{equation}
\langle \xi _{i}(t_{1})\xi _{j}(t_{2})\rangle =2\Gamma _{0}^{-1}\delta
_{i,j}\,\delta (t_{1}-t_{2})
\end{equation}
Below we shall choose the time units so that $\Gamma _{0}=1$. Using the
standard approach \cite{Fischer93} we get a path integral formulation of the
problem with the Lagrangian 
\begin{equation}
\mathcal{L}=\sum_{i}-\hat{\sigma}_{i}^{2}-i\hat{\sigma}_{i}\left( \partial
_{t}\sigma _{i}+{\frac{{\delta \beta \mathcal{H}}}{{\delta \sigma _{i}}}}%
\right) +{\frac{1}{2}}{\frac{{\delta ^{2}\beta \mathcal{H}}}{{\delta \sigma
_{i}^{2}}}}.  \label{lagr}
\end{equation}
The Green functions are defined by 
\begin{eqnarray}
\mathcal{G}_{i,j}(t_{1},t_{2}) &\equiv &\left[ 
\begin{array}{cc}
\hat{D}_{i,j}(t_{1},t_{2}) & G_{i,j}^{\dagger }(t_{1},t_{2}) \\ 
G_{i,j}(t_{1},t_{2}) & D_{i,j}(t_{1},t_{2})
\end{array}
\right]  \nonumber \\
&=&\int D\sigma D\hat{\sigma}\,\,e^{\int_{t_{i}}^{t_{f}}\mathcal{L}%
\,dt}\,\,\left[ 
\begin{array}{c}
i\hat{\sigma}_{i}(t_{1}) \\ 
\sigma _{i}(t_{1})
\end{array}
\right] [i\hat{\sigma}_{j}(t_{2}),\sigma _{j}(t_{2})],
\end{eqnarray}
 and the dynamical action $A$ is defined by 
\begin{equation}
e^{A}=\int D\sigma D\hat{\sigma}e^{\int_{t_{i}}^{t_{f}}\mathcal{L}\,dt}.
\label{prob}
\end{equation}

Usually one fixes only initial boundary conditions. In this case the Green
function G is casual, the anomalous Green function $\hat{D}$ is zero, and
the action vanishes. But if one considers rare processes as transitions over
the barriers, then one should also fix the final boundary conditions. In
that case the Green function G does not need to be causal, and the action
and the Green function $\hat{D}$ do not necessarily vanish.

Now we construct a mapping between an uphill motion (with a negative action
monotonically decreasing in time) and the downhill motion (with zero action)
going back in time. Consider arbitrary instanton process ($\sigma (t),%
\widehat{\sigma }(t)$) and impose both initial and final boundary
conditions. Applying the transformation

\begin{equation}
\lbrack i\hat{\sigma},\sigma ]\to [i\hat{\sigma}+\vec{\partial}_{t}\sigma
,\sigma ]  \label{trsigma}
\end{equation}
to the Lagrangian (\ref{lagr}) we get

\begin{eqnarray}
\mathcal{L} &\to &\mathcal{L}_{n}=\sum_{i}-\hat{\sigma}_{i}^{2}-i\hat{\sigma}%
_{i}\left( -\partial _{t}\sigma _{i}+{\frac{{\delta \beta \mathcal{H}}}{{%
\delta \sigma _{i}}}}\right) +{\frac{1}{2}}{\frac{{\delta ^{2}\beta \mathcal{%
H}}}{{\delta \sigma _{i}^{2}}}}  \nonumber \\
&&+\beta \left( H(t_{f})-H(t_{i})\right)  \label{lagr1}
\end{eqnarray}
Note that this Lagrangian differs from the original Lagrangian (\ref{lagr})
only by the sign of time derivative and constant boundary term. Therefore
inverting time in the Lagrangian (\ref{lagr1}) one can make the Lagrangians (%
\ref{lagr},\ref{lagr1}) equivalent. Therefore, the Lagrangian (\ref{lagr1})
describes normal downhill motion formally inverted in time.

The Green function defined as averaged with respect to Lagrangian (3) 
\begin{equation}
\mathcal{G}(t_{1},t_{2})=\left\langle \left[ 
\begin{array}{c}
i\hat{\sigma}(t_{1}) \\ 
\sigma (t_{1})
\end{array}
\right] [i\hat{\sigma}(t_{2}),\sigma (t_{2})]\right\rangle _{\mathcal{L}}
\end{equation}
and the Green function defined as averaged with respect to the Lagrangian (%
\ref{lagr1}) 
\begin{equation}
\mathcal{G}_{n}(t_{1},t_{2})=\left\langle \left[ 
\begin{array}{c}
i\hat{\sigma}(t_{1}) \\ 
\sigma (t_{1})
\end{array}
\right] [i\hat{\sigma}(t_{2}),\sigma (t_{2})]\right\rangle _{\mathcal{L}_{n}}
\end{equation}
are related by 
\begin{equation}
\mathcal{G}(t_{1},t_{2})=\left[ 
\begin{array}{cc}
1 & \overrightarrow{\partial }_{t1} \\ 
0 & 1
\end{array}
\right] \mathcal{G}_{n}(t_{1},t_{2})\left[ 
\begin{array}{cc}
1 & 0 \\ 
\overleftarrow{\partial }_{t2} & 1
\end{array}
\right] .  \label{transform}
\end{equation}
Because the Green function $\mathcal{G}_{n}$ corresponds to the normal
downhill process inverted in time, it should have the form 
\begin{equation}
\mathcal{G}_{n}=\left[ 
\begin{array}{cc}
0 & G_{n}^{\dagger } \\ 
G_{n} & D_{n}
\end{array}
\right] .
\end{equation}
Therefore the transformation (\ref{transform}) in components is

\begin{equation}
D(t_1,t_2)=D_n(t_1,t_2)  \label{trcom4}
\end{equation}
\begin{equation}
G(t_1,t_2)=G_n(t_1,t_2)+\partial_2 D_n(t_1,t_2)  \label{trcom2}
\end{equation}
\begin{equation}
\hat D(t_1,t_2)=\partial_1 G_n(t_1,t_2)+\partial_2 G_n^\dagger(t_1,t_2)
+\partial_1\partial_2 D_n(t_1,t_2).  \label{trcom1}
\end{equation}
Note that the response function $G_n$ should be purely advanced due to
formal inversion of time with respect to the downhill motion.

Thus, in order to construct the Green functions for the instanton processes
one should find the Green functions for the corresponding normal process,
invert time, and apply the transformation (\ref{transform}).

\begin{figure}[h]
\unitlength1.0cm
\par
\begin{center}
\begin{picture}(6,4)
\epsfxsize=6.0 cm
\epsfysize=4.0 cm
\epsfbox{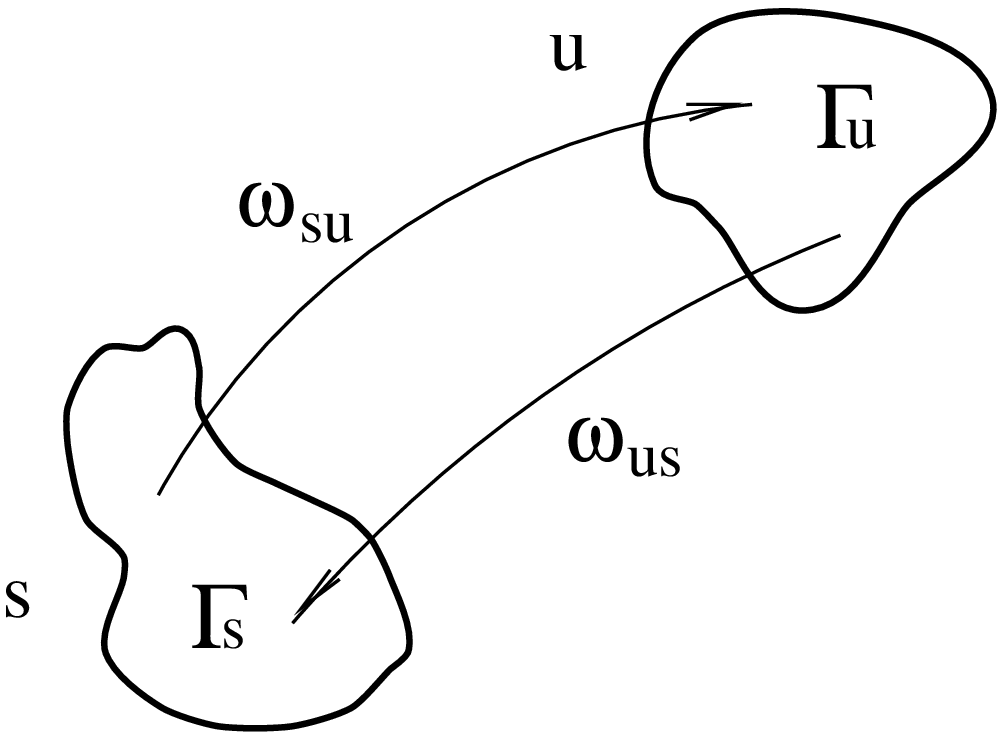}
\end{picture}
\end{center}
\caption{}
\end{figure}

Now we show that the action can be expressed through the energies and
configuration spaces of the initial and final states: Suppose that initially
the system is at the stable state $s$, and we want to find the probability
to escape this state going trough the unstable fixed point $u$ which is the
final state of the instanton motion. Note that in all the above
considerations we assumed completely fixed boundary conditions for $\sigma
_{i}.$ But physically, initial and final states correspond to some regions
in configuration states which we denote as $\Gamma _{s}$ and $\Gamma _{u}$
respectively. To emphasize the difference between the process with
completely fixed boundary conditions and the processes with physical
boundary conditions we refer to the former as elementary processes.
According to Eqs.(\ref{lagr},\ref{lagr1}) the probability of an elementary
process of motion from $s$ to $u$ ($w_{su}$) is related to the elementary
process of motion from $u$ to $s$ ($w_{us}$) through

\begin{equation}
w_{su}=w_{us}\,e^{\beta (E(t_{f})-E(t_{i}))}.  \label{wrel}
\end{equation}
The probability to escape the state $s$ is 
\begin{equation}
W_{su}=\Gamma _{u}w_{su}.  \label{Wsu}
\end{equation}
On the other hand the probability to go from the unstable state $u$ to the
stable state $s$ is 1, therefore 
\begin{equation}
W_{us}=w_{us}\Gamma _{s}=1.  \label{Wus}
\end{equation}
Combining Eqs.(\ref{wrel},\ref{Wsu},\ref{Wus}) for the probability to escape
the stable state $u$ we get 
\begin{equation}
W_{su}=\exp [-\beta (E(t_{f})-E(t_{i}))+S(t_{f})-S(t_{i})],  \label{prob1}
\end{equation}
where $S=\ln \Gamma $ is the entropy. Note that at the stable point the
entropy $S(t_{i})$ is just the equilibrium entropy corresponding to this
state. The entropy of the final state $S(t_{f})$ can not be defined
thermodynamically because it corresponds to the configuration space of the
unstable state. Defining the free energy as $F=E-TS$ one can write ($\ref
{prob1}$) as 
\begin{equation}
W_{su}=e^{-\beta F(t_{f})+\beta F(t_{i})}.  \label{actiong}
\end{equation}

\section{Instanton transition in the spherical SK model.}

\label{3}

The Hamiltonian of the spherical SK model is 
\begin{equation}
\mathcal{H}={\frac{1}{2}}\sum \sigma _{i}J_{ij}\sigma _{j},
\end{equation}
with the spherical constraint $\sum_{i}\sigma _{i}\sigma _{i}=N$ imposed.
The Hamiltonian becomes diagonal in the basis of eigenvectors of the matrix $%
J_{i,j}$ 
\begin{equation}
\mathcal{H}={\frac{1}{2}}\sum_{\mu }\epsilon _{\mu }\,s_{\mu }^{2},
\end{equation}
where 
\begin{eqnarray}
\sum_{j}J_{i,j}\sigma _{j}^{\mu } &=&\epsilon _{\mu }\sigma _{i}^{\mu }, \\
\sigma _{i} &=&\sum_{\mu }s_{\mu }\,\sigma _{i}^{\mu }.
\end{eqnarray}
The Lagrangian corresponding to this model is 
\begin{equation}
\mathcal{L}(s,\lambda )=-\sum_{\mu }\left[ \hat{s}_{\mu }^{2}+i\hat{s}_{\mu
}(\partial _{t}+\epsilon _{\mu }+\lambda )s_{\mu }\right] +{\frac{1}{2}}%
(N+2)\lambda ,
\end{equation}
where $\lambda $ is the time-dependent Lagrange multiplier field which
appears in the equation of motion due to the constraint and we take $\beta
=1 $ for convenience. The functional integral over the variables $s_{\mu },%
\widehat{s}_{\mu }$ should be performed with the weight that includes $%
\delta -$function that ensures the constraint. Using the integral
representation of this $\delta -$function 
\begin{equation}
\delta (\sum_{\mu }s_{\mu }^{2}-N)=\int D\phi \,e^{i\int dt\,\,\phi
\,\,(\sum_{\mu }s_{\mu }^{2}-N)},
\end{equation}
we get the following Lagrangian 
\begin{eqnarray}
\mathcal{L}(s,\hat{s},\lambda ,\phi ) &=&\sum_{\mu }\left( -\hat{s}_{\mu
}^{2}-i\hat{s}_{\mu }(\partial _{t}+\epsilon _{\mu }+\lambda )s_{\mu }+i\phi
\,\,s_{\mu }^{2}\right)  \nonumber \\
&&+{\frac{1}{2}}(N+2)\lambda -i\phi N.
\end{eqnarray}
At low temperatures the condensation into the lowest eigenvalue $\mu =0$
eventually takes place. Therefore we introduce the condensate $S_{0}$ and
integrate over $s_{\mu }$ with $\mu \ge 1$ getting 
\[
\mathcal{L}=-\hat{S}_{0}^{2}-i\hat{S}_{0}(\partial +\epsilon _{0}+\lambda
)S_{0}+i\phi \,\,S_{0}^{2}+{\frac{1}{2}}(N+2)\lambda -i\phi N 
\]
\begin{equation}
-{\frac{1}{2}}\sum_{\mu \ge 1}Tr\ln \mathcal{G}_{\mu },  \label{splag}
\end{equation}
where the matrix Green function 
\begin{equation}
\mathcal{G}_{\mu }=\left[ 
\begin{array}{cc}
\hat{D}_{\mu }(t_{1},t_{2}) & G_{\mu }^{\dagger }(t_{1},t_{2}) \\ 
G_{\mu }(t_{1},t_{2}) & D_{\mu }(t_{1},t_{2})
\end{array}
\right]
\end{equation}
satisfies the equation

\begin{equation}
\left[ 
\begin{array}{cc}
-2 & \partial_{t1}+\lambda(t_1)+\epsilon_\mu \\ 
-\partial_{t1}+\lambda(t_1)+\epsilon_\mu & \phi(t_1)
\end{array}
\right] \mathcal{G}_\mu=\delta(t_1-t_2).  \label{spgrf}
\end{equation}

The number of cites $N$ is large, therefore we will perform the integrals
with the weight $\exp (\int \mathcal{L}dt)$ where $\mathcal{L}$ is given by (%
\ref{splag}) in the saddle point approximation. Taking the variation with
respect to $\phi ,\lambda ,\hat{S}_{0},S_{0}$ and transforming the variables
via $-2i\phi \to \phi ,\,\,\,i\hat{S}\to \sqrt{N}\hat{S}$ we get the saddle
point equations 
\begin{eqnarray}
{\frac{1}{N}}\sum_{\mu \ge 1}D_{\mu }+S_{0}^{2} &=&1,  \label{sadlpf1} \\
{\frac{1}{N}}\sum_{\mu \ge 1}G_{\mu }+\hat{S}_{0}S_{0} &=&0,  \label{sadlpf2}
\\
(\partial +\epsilon _{0}+\lambda )S_{0}-2\hat{S}_{0} &=&0,  \label{sadlpf3}
\\
(-\partial +\epsilon _{0}+\lambda )\hat{S}_{0}+\phi S_{0} &=&0,
\label{sadlpf4}
\end{eqnarray}
where 
\[
D_{\mu }(t)=D_{\mu }(t,t),\,\,\,\,G_{\mu }(t)=G_{\mu }(t,t+\delta ). 
\]
Note that Eqs.(\ref{sadlpf1}-\ref{sadlpf4}) contain only the equal time
Green functions. Therefore it is convenient to write the equations directly
for the equal time Green functions instead of (\ref{spgrf}): 
\begin{eqnarray}
\partial G_{\mu } &=&2\hat{D}_{\mu }+\phi D_{\mu },  \label{sadlpf5} \\
\partial \hat{D}_{\mu } &=&2(\lambda +\epsilon _{\mu })\hat{D}_{\mu }+\phi
\,(1+2G_{\mu }),  \label{sadlpf6} \\
\partial D_{\mu } &=&-2(\lambda +\epsilon _{\mu })D_{\mu }+2(1+2G_{\mu }),
\label{sadlpf7}
\end{eqnarray}
We assume that the system is initially at the equilibrium. This corresponds
to the stationary solution of Eqs.(\ref{sadlpf1}-\ref{sadlpf7}) 
\begin{eqnarray}
{\frac{1}{N}}\sum_{\mu \ge 1}D_{\mu }+S_{0}^{2} =1,  \label{stp1} \\
D_{\mu } ={\frac{1}{{\lambda _{s}+\epsilon _{\mu }}}},  \label{stp2} \\
\lambda _{s} =-\epsilon _{0},  \label{stp3} \\
G_{\mu } =\hat{D}_{\mu }=\hat{S}_{0}=\phi =0,
\end{eqnarray}
where $\lambda _{s}$ is the value of $\lambda $ for this stable stationary
solution. Eq.(\ref{stp3}) follows from Eq.(\ref{sadlpf3}) because we assumed
that there is a nonzero condensate density $S_{0}.$ Note that there are two
equilibrium states with $S_{0}>0$ and $S_{0}<0.$

Our goal is to find the probability of the instanton transition from one
state to the other. For definiteness, assume that the initial state is one
with $S_{0}>0.$ Obviously, $S_{0}$ first decreases to zero during the
instanton process and then it becomes negative. Only the first uphill part
of the motion gives the contribution to the action, therefore we will
consider only this uphill part of the trajectory. The end point of this
uphill motion corresponds to an unstable fixed point of Eqs.(\ref{sadlpf1}-%
\ref{sadlpf7}). The condensate density $S_{0}$ is zero at this point
therefore from Eqs.(\ref{sadlpf1}-\ref{sadlpf7}) we get the following
equations corresponding to this stationary (but unstable) solution: 
\begin{eqnarray}
{\frac{1}{N}}\sum_{\mu \ge 1}D_{\mu } &=&1, \\
D_{\mu } &=&{\frac{1}{{\lambda _{u}+\epsilon _{\mu }}}}, \\
G_{\mu } &=&\hat{D}_{\mu }=\hat{S}_{0}=\phi =0,
\end{eqnarray}
where $\lambda _{u}$ is the value of $\lambda $ at this unstable stationary
solution. Physically, it is natural to expect that that this solution
corresponds to the condensate in the first eigenstate $\mu =1.$ Indeed,
taking $\lambda _{u}=-\epsilon _{1}+{\frac{1}{{S_{1}^{2}}}},$ where $S_{1}$
is the condensate at the first eigenstate $\mu =1,$ we get 
\begin{eqnarray}
{\frac{1}{N}}\sum_{\mu \ge 2}D_{\mu }+S_{1}^{2} &=&1, \\
D_{\mu } &=&{\frac{1}{{\lambda _{u}+\epsilon _{\mu }}}}.
\end{eqnarray}
These equations are similar to Eqs.(\ref{stp1}-\ref{stp3}) for the stable
fixed point with the difference that the system condenses into the first
eigenstate.

Now we need to find the trajectory connecting the fixed points mentioned
above. It was shown in the previous section that an uphill trajectory can be
mapped into a downhill trajectory going back in time. To show this,
according to Eqs.(\ref{trsigma},\ref{trcom2}), one should take 
\begin{eqnarray}
G_{\mu } &=&{\frac{1}{2}}\partial D_{\mu },  \label{sptr1} \\
\hat{S}_{0} &=&\partial S_{0},  \label{sptr2}
\end{eqnarray}
where $D_{\mu },S_{0}$ should satisfy the downhill equations with inverse
time 
\begin{eqnarray}
{\frac{1}{N}}\sum_{\mu \ge 1}D_{\mu }+S_{0}^{2} &=&1,  \label{spdown1} \\
(-\partial +\epsilon _{0}+\lambda )S_{0} &=&0,  \label{spdown2} \\
-\partial D_{\mu } &=&-2(\lambda +\epsilon _{\mu })D_{\mu }+2
\label{spdown3}
\end{eqnarray}
Indeed, taking 
\begin{eqnarray}
\phi &=&\partial \lambda , \\
\hat{D}_{\mu } &=&{\frac{1}{4}}\partial ^{2}D_{\mu }-{\frac{1}{2}}D_{\mu
}\partial \lambda ,
\end{eqnarray}
along with Eqs.(\ref{sptr1},\ref{sptr2}), one can show that Eqs.(\ref
{sadlpf1}-\ref{sadlpf7}) are reduced to Eqs.(\ref{spdown1}-\ref{spdown3}).
The trajectory connecting the stable and unstable saddle points can be found
analytically; (see Appendix A) the result is 
\begin{eqnarray}
\lambda (t)+\epsilon _{0} &=&(\lambda _{u}+\epsilon _{0})f[2\,|\lambda
_{u}+\epsilon _{0}|\,(t_{0}-t)] \\
D_{\mu }(t) &=&{\frac{1}{{\lambda _{u}+\epsilon _{\mu }}}}f[2\,|\lambda
_{u}+\epsilon _{0}|\,(t_{0}-t)]  \nonumber \\
&&+{\frac{1}{{\epsilon _{\mu }}}}f[2|\lambda _{u}+\epsilon _{0}|(t-t_{0})],
\end{eqnarray}
where 
\begin{equation}
f[x]={\frac{1}{{e^{x}+1}}},
\end{equation}
where $t_{0}$ is an arbitrary finite time which reflects the translational
invariance in time.

The last step is to calculate the action which determines the probability of
the instanton process. Note, that to find it, one should take $Tr\ln $ of
the operator containing $\lambda ,\phi $ which are functions of time. The
necessary calculation is presented in Appendix B, and the answer, simplified
with the help of Eqs.(\ref{sadlpf1}-\ref{sadlpf7}), is 
\begin{equation}
A/N={\frac{1}{2}}\int dt\left( \phi +{\frac{2}{N}}\sum_{\mu \ge 1}{\frac{{%
G_{\mu }}}{{D_{\mu }}}}\right) .
\end{equation}
Using that $\phi =\partial \lambda $ and $G_{\mu }={\frac{1}{2}}\partial
D_{\mu }$ we get 
\begin{equation}
A/N={\frac{1}{2}}\left( \lambda +{\frac{1}{N}}\sum_{\mu \ge 1}\ln D_{\mu
}\right) _{t=t_{i}}^{t=t_{f}}.
\end{equation}
Note that at the fixed points $D_{\mu }=1/(\epsilon _{\mu }+\lambda ),$
therefore one can write 
\begin{equation}
A=-\Bigl[F(\lambda (t_{f}))-F(\lambda (t_{i}))\Bigr],  \label{delf}
\end{equation}
where $F(\lambda )$ is the equilibrium free energy of this model 
\begin{equation}
F(\lambda )/N=-\lambda /2+{\frac{1}{{2N}}}\sum_{\mu \ge 1}\ln (\epsilon
_{\mu }+\lambda ).  \label{fener}
\end{equation}
The result (\ref{delf}) is in the agreement with the general result (\ref
{actiong}). As we mentioned in Sec.\ref{2} the free energy at the unstable
fixed point cannot be defined thermodynamically. But in this simple model
one can formally eliminate the unstable direction (i.e. impose the
constraint $S_0 =0$) and then it becomes possible to define the free energy
thermodynamically. That is why we got the difference of the equilibrium free
energies in (\ref{fener}). Note that the Lagrange multiplier $\lambda $
changes during the instanton process on $\lambda (t_{f})-\lambda
(t_{i})=\epsilon _{1}-\epsilon _{0}.$ The typical distance between
neighboring energy levels at the edges of energy level distribution is of
order of $1/\sqrt{N}.$ Therefore one can expand (\ref{fener}) in $\Delta
\lambda $ getting 
\begin{equation}
A=-{\frac{N}{2T}}S_{0}^{2}(\epsilon _{1}-\epsilon _{0}),
\end{equation}
where we restored the temperature $T$. Note that this action is of the order
of $\sqrt{N}.$

\section{Discussion and conclusions.}

\label{4}

We developed the method that simplifies the calculation of the probablity of
rare processes such as transitions over the barriers. Our method is based on
the Lagrangian approach to the dynamics in which rare processes correspond
to the instantons. Generally, in order to obtain the probability of a
particular transition between two given states one need to apply a boundary
condition in future and in the past, this destroys the causality of the
theory: the response function $G=\langle \hat{s}s\rangle $ becomes non
causal and the anomalous correlation function $\hat{D}=\langle \hat{s}\hat{s}%
\rangle $ appears. This complicates the description of the instanton
processes.

The main result of this paper is that an instanton process can be mapped
into a usual process going back in time. So, knowing the correlation
functions of the corresponding normal process one can construct the
instanton correlation functions. We showed that this mapping gives the
sensible probability to escape a free energy well, $e^{-\Delta F/T}$, where $%
\Delta F$ is the depth of the free energy well. The free energy at the end
of the instanton trajectory cannot be defined thermodynamically because at
this point the system is at the unstable equilibrium, instead it should be
defined by $F=E-TS$, where $E$ is the energy and $S$ is the statistical
entropy at the end of the instanton trajectory, i.e. entropy constrained to
the states orthogonal to the descending direction.

We applied this approach to the spherical SK model which usual dynamical
properties were studied in Ref.\cite{cug}. This model has just two ground
states, corresponding to the energy level $\epsilon _{0}$, and no other
metastable states. Although the relaxation towards each of these two states
is exponential, model exhibits aging behavior when the system relaxes from a
random spin configuration to the equilibrium. We considered the instanton
transition from one ground state to the other. In accordance with our
general result the equations describing the instanton process in this model
can be transformed into the usual equations with inverted time. This
transformation allows one to find analytically the instanton trajectory. The
probability of this transition was found to be $e^{-S_{0}^{2}N(\epsilon
_{1}-\epsilon _{0})/2T}$ where $\epsilon _{1}$ is the energy of the first
(unstable) level. It shows that although the system has $N$ saddle points
and a complicated phase space the path connecting two ground states with
opposite magnetization might go via the saddle point with the lowest energy.
The typical distance between the neighboring energy levels at the edge of
the energy spectrum is of order $1/\sqrt{N},$ therefore the action is of
order $\sqrt{N}.$ Note that the distance between the energy levels $\epsilon
_{1}-\epsilon _{0}$ is different for different samples, therefore the
transition probability is not a self-averaging quantity. Therefore in this
problem it would be very difficult to get the correct answer for the
transition probability in any technique which involves averaging at the
beginning of calculations.

We hope that this method can be used to find the barriers between the
metastable states in more complicated spin glasses like $p>2$ spin models or
SK model. The first attempt of application of the instanton method to SK
model was done in Ref.\cite{Ioffe98} It is a more complicated problem because
there are many metastable states in these glasses and therefore the
averaging should be done at the beginning of calculations. The scaling of
the action $\sqrt{N}$ which we got for $p=2$ model  is probably specific for
the spherical model because the barriers in spin glasses with exponential
number of states are due to the nonlinearity of the dynamical equations
which is absent in $p=2$ spherical model.

\section*{Appendix A}

In this Appendix we will find the trajectory connecting the unstable
fixed point with the stable one. It is natural to invert time in Eqs.(\ref
{spdown1}-\ref{spdown3}) so that they will describe the usual downhill
motion: 
\begin{eqnarray}
{\frac{1}{N}}\sum_{\mu }\tilde{D}_{\mu }+\tilde{S}_{0}^{2} &=&1,
\label{downmot1} \\
(\partial +\epsilon _{0}+\tilde{\lambda})\tilde{S}_{0} &=&0,
\label{downmot2} \\
\partial \tilde{D}_{\mu } &=&-2(\tilde{\lambda}+\epsilon _{\mu })\tilde{D}%
_{\mu }+2,  \label{downmot3}
\end{eqnarray}
where tilde means invertion of time with respect to the instanton motion,
for example $\tilde{S}_{0}(t)=S_{0}(-t).$ We need to find the solution of
these equatiosn corresponding to the downhill trajectory that begins from
the unstable fixed point and ends at the stable one. Therefore the initial
boundary condition is ($t=-\infty $) 
\begin{eqnarray}
{\tilde D}_{\mu }(-\infty ) &=&{\frac{1}{{\epsilon _{\mu }+\lambda_{u}%
}}},\,\,\,\,\mu \ge 1,  \label{bcon} \\
S_{0} &=&0, \\
\tilde{\lambda}(-\infty ) &=&-\epsilon _{1}+{\frac{1}{{\ S_{1}^{2}}}}, \\
\tilde{\lambda}(-\infty ) &=&\lambda _{u},
\end{eqnarray}
and the final one ($t=\infty $) is 
\begin{eqnarray}
\tilde{D}_{\mu }(\infty ) &=&{\frac{1}{{\epsilon _{\mu }-\epsilon _{0}}}}%
,\,\,\,\,\mu \ge 1, \\
\tilde{\lambda}(\infty ) &=&-\epsilon _{0}, \\
S_{0} &\ne &0.
\end{eqnarray}
The solution of the Eq.(\ref{downmot3}) satisfying to the boundary condition
(\ref{bcon}) is 
\begin{equation}
\tilde{D}_{\mu }(t)=2\int_{-\infty }^{t}dt^{\prime }e^{-2\int_{t^{\prime
}}^{t}[\tilde{\lambda}(t^{\prime \prime })+\epsilon _{\mu }]dt^{\prime
\prime }}  \label{atd}
\end{equation}
Using Eqs.(\ref{downmot2},\ref{downmot3}) one can write Eq.(\ref{downmot1})
in the form 
\begin{equation}
\tilde{\lambda}+\epsilon _{0}=-{\frac{1}{N}}\sum_{\mu }(\epsilon _{\mu
}-\epsilon _{0})D_{\mu }+1,  \label{alam}
\end{equation}
which will be more convenient for us. For simplisity let us take $\epsilon
_{0}=0$ further in this Appendix. Substitution of Eq.(\ref{atd}) into Eq.(%
\ref{alam}) gives 
\begin{equation}
\tilde{\lambda}(t)={\frac{2}{N}}\sum_{\mu }\int_{-\infty }^{t}dt^{\prime
}e^{-2\int_{t^{\prime }}^{t}[\tilde{\lambda}(t^{\prime \prime })+\epsilon
_{\mu }]dt^{\prime \prime }}\tilde{\lambda}(t^{\prime }).
\end{equation}
Introducing the function 
\begin{equation}
F(t)=\Gamma (t)\tilde{\lambda}(t),  \label{agamma}
\end{equation}
where $\Gamma (t)$ is defined by (up to the multiplication by a constant) 
\begin{equation}
{\frac{{\Gamma (t)}}{{\Gamma (t^{\prime })}}}=e^{2\int_{t^{\prime }}^{t}%
\tilde{\lambda}(t^{\prime \prime })dt^{\prime \prime }},
\end{equation}
one can write Eq.(\ref{alam}) as a linear integral equation on $F(t)$ 
\begin{equation}
F(t)={\frac{2}{N}}\int_{-\infty }^{t}dt^{\prime }F(t^{\prime })\sum_{\mu
}e^{-2\epsilon _{\mu }(t-t^{\prime })},
\end{equation}
which can be easily solved giving 
\begin{equation}
F(t)=A\,\,e^{2\lambda _{u}t},
\end{equation}
where $A$ is an arbitrary constant. Now using Eq.(\ref{agamma}) one can
write 
\begin{equation}
e^{2\lambda _{u}(t-t^{\prime })}={\frac{{\lambda (t)}}{{\lambda (t^{\prime })%
}}}e^{2\int_{t^{\prime }}^{t}\tilde{\lambda}(t^{\prime \prime })dt^{\prime
\prime }},
\end{equation}
then taking a logarithm and differentiating with respect to $t$ we get
a differential equation on $\tilde{\lambda}$ 
\begin{equation}
\partial_t \tilde{\lambda}(t)=2\tilde{\lambda}(t)[\tilde{\lambda}_{u}-\tilde{%
\lambda}(t)],
\end{equation}
which can be solved giving 
\begin{equation}
\tilde{\lambda}(t)=\tilde{\lambda}_{u}f[2\,|\tilde{\lambda}_{u}|\,(t-t_{0})],
\label{sollam}
\end{equation}
where 
\begin{equation}
f[x]={\frac{1}{{e^{x}+1}}},
\end{equation}
and $t_{0}$ is an arbitrary time which reflects the translational invariance
in time. In case $\epsilon _{0}\ne 0$ one can easily generalize Eq.(\ref
{sollam}) to 
\begin{equation}
\tilde{\lambda}(t)+\epsilon _{0}=(\tilde{\lambda}_{u}+\epsilon _{0})f[2\,|%
\tilde{\lambda}_{u}+\epsilon _{0}|\,(t-t_{0})].
\end{equation}
Knowing $\tilde{\lambda}(t)$ one can find $\tilde{D}_{\mu }$ 
\begin{eqnarray}
\tilde{D}_{\mu }(t) &=&{\frac{1}{{\lambda _{u}+\epsilon _{\mu }}}}%
f[2\,|\lambda _{u}+\epsilon _{0}|\,(t-t_{0})]  \nonumber \\
&&+{\frac{1}{{\epsilon _{\mu }}}}f[2\,|\lambda _{u}+\epsilon
_{0}|\,(t_{0}-t)].
\end{eqnarray}

\section*{Appendix B}

The main problem in calculation of the action is to find 
\begin{equation}
A_\mu\equiv{\frac{1}{2}}Tr \ln \left[ 
\begin{array}{cc}
-2 & \partial+\lambda+\epsilon_\mu \\ 
-\partial+\lambda+\epsilon_\mu & \phi
\end{array}
\right].  \label{appb1}
\end{equation}
Note that taking the variational derivative of (\ref{appb1}) with respect to 
$\lambda(t)$ and $\phi(t)$ we get respectively $G_\mu(t,t)=1/2+G_\mu(t)$ and 
${\frac{1}{2}} D_\mu (t).$ The idea of our method of calculation of (\ref
{appb1}) is to find a functional which gives the same functions ($G_\mu+{%
\frac{1}{2}}$ and ${\frac{1}{2}} D_\mu$) when one takes the variations with
respect to $\lambda$ and $\phi.$ Up to boundary terms this functional should
be equal to the action, and these boundary terms can be found from the
requirement that the action should be zero for any downhill trajectory.

Now let us find the functional mentioned above: Eqs.(\ref{sadlpf5}-\ref
{sadlpf7}) have the following invariant 
\begin{equation}
D_\mu\hat D_\mu-(1+G_\mu)G_\mu=c,  \label{appb2}
\end{equation}
where $c$ is an arbitrary constant. But initially $G_\mu=\hat D_\mu=0,$
therefore $c=0.$ The condition (\ref{appb2}) can be satisfied automatically
by introducing the new variables 
\begin{equation}
D_\mu=\eta_\mu\eta_\mu, \,\,\,\, G_\mu=\eta_\mu\hat \eta_\mu-1/2, \,\,\,\,
\hat D_\mu=\hat \eta_\mu\hat \eta_\mu-{\frac{1}{{4 \eta_\mu^2}}}.
\end{equation}
The new variables $\eta_\mu,\hat\eta_\mu$ satisfy the following equations 
\begin{eqnarray}
\partial\eta_\mu=-(\lambda+\epsilon_\mu)\eta_\mu+2\hat\eta_\mu,
\label{appbe1} \\
\partial\hat\eta_\mu=(\lambda+\epsilon_\mu)\hat\eta_\mu+\phi\,\eta_\mu -{%
\frac{1}{{2 \eta_\mu^3}}} .  \label{appbe2}
\end{eqnarray}
These equations can be obtained by taking the variations of the functional 
\begin{equation}
\Gamma_\mu=-\left[\hat\eta_\mu^2-\hat\eta_\mu(\partial+\epsilon_\mu+\lambda)%
\eta_\mu -{\frac{1}{2}}\phi\,\eta^2_\mu-{\frac{1}{{4\eta_\mu^2}}}\right]
\end{equation}
with respect to $\hat\eta$ and $\eta.$ Note that the variational derivatives
of (\ref{appb1}) and $\Gamma_\mu$ with respect to $\lambda,\phi$ are the
same if we take $\Gamma_\mu$ at the saddle point with respect to 
 $\hat\eta,\eta,$ i.e. Eqs.(\ref{appbe1},\ref{appbe2}) should be
satisfied.  Therefore
the action $A_\mu$ can be written as 
\begin{equation}
A_\mu=\Gamma_\mu+T\left.\Bigl(\hat
D_\mu(t),D_\mu(t),G_\mu(t),\lambda(t),\phi(t)\Bigr) \right|_{t=t_i}^{t=t_f},
\label{appb3}
\end{equation}
where $T$ is an unknown function. The initial and final times $t_i$ and $t_f 
$ correspond to the fixed points of Eqs.(\ref{sadlpf1}-\ref{sadlpf7}),
therefore the abnormal functions $\hat D_\mu,G_\mu,\phi$ should be zero at
these points and we can simplify (\ref{appb3}) 
\begin{equation}
A_\mu=\Gamma_\mu+T\left.\Bigl(D_\mu(t),\lambda(t)\Bigr)
\right|_{t=t_i}^{t=t_f}.
\end{equation}
Using the saddle point equations (\ref{sadlpf1}-\ref{sadlpf7}) and Eqs.(\ref
{appbe1},\ref{appbe2}) one can simplify the total action $A$ getting 
\begin{eqnarray}
A/N={\frac{1}{2}}\int dt\left(\phi+{\frac{1}{N}}\sum_\mu{\frac{{2G_\mu}}{{%
D_\mu}}} -{\frac{1}{{2N}}}\sum_\mu\partial \ln D_\mu\right)  \nonumber \\
+\sum_\mu T\left.\Bigl(D_\mu(t),\lambda(t)\Bigr) \right|_{t=t_i}^{t=t_f}.
\label{appb5}
\end{eqnarray}
But this action should be zero for any downhill trajectory, therefore the
third and forth terms in (\ref{appb5}) should cancel each other, and we
finally get 
\begin{equation}
A/N={\frac{1}{2}}\int dt\left(\phi+{\frac{1}{N}}\sum_\mu{\frac{{2G_\mu}}{{%
D_\mu}}}\right).
\end{equation}


\begin{thebibliography}{9}
\bibitem{Fischer93}  (see for example) K.H. Fischer and J.A. Hertz, \textit{%
Spin Glasses} (Cambridge University Press, 1993)

\bibitem{cgp}  A.Cavagna, I. Giardina, G. Parisi, cond-mat/9702069

\bibitem{kucu}  L.F. Cugliandolo and J. Kurchan, Phys. Rev. Let. B \textbf{71%
}, 173 (1993)

\bibitem{pspin}  A. Crisanti, H. Horner, H.-J. Sommers, Z. Phys. B \textbf{92%
}, 257 (1993)

\bibitem{kirk}  T. R. Kirkpatrick, D. Thirumalai, Phys. Rev. B \textbf{36 },
5388 (1987)

\bibitem{Dotsenko90}  V. Dotsenko, M. Feigelman and L. B. Ioffe, \emph{Spin
Glasses and related problems, } Soviet Physics Reviews  \textbf{A15}, 1-250
(1990); L. B. Ioffe, Phys. Rev. B \textbf{38}, 5181 (1988).

\bibitem{cug}  Leticia F. Cugliandolo and David S. Dean, J. Phys. A \textbf{%
28}, 4213 (1995)

\bibitem{Ioffe98}  L.B. Ioffe, D. Sherrington, Phys. Rev. Rev. B \textbf{57},
7666 (1998)
\end{thebibliography}
\end{document}